\documentclass[aps,prd,twocolumn,showpacs,floatfix,superscriptaddress,
preprintnumbers,nofootinbib]{revtex4}

\usepackage{epsfig,amssymb}

\usepackage[dvips]{color}
\definecolor{Black}{named}{Black}
\definecolor{Red}{named}{Red}
\definecolor{Blue}{named}{Blue}

\def\lsim{\raise0.3ex\hbox{$\;<$\kern-0.75em\raise-1.1ex\hbox{$\sim\;$}}}
\def\gsim{\raise0.3ex\hbox{$\;>$\kern-0.75em\raise-1.1ex\hbox{$\sim\;$}}}

\def\theta{\vartheta}

\newcommand{\be}{\begin{equation}}
\newcommand{\ee}{\end{equation}}
\newcommand{\ba}{\begin{eqnarray}}
\newcommand{\ea}{\end{eqnarray}}

\begin{document}


\title{Blazar halos as probe for extragalactic magnetic fields and
maximal acceleration energy}

\author{K.~Dolag}
\affiliation{Max-Planck-Institut f\"ur Astrophysik, Garching, Germany}

\author{M.~Kachelrie\ss}
\affiliation{Institutt for fysikk, NTNU, Trondheim, Norway}

\author{S.~Ostapchenko}
\affiliation{Institutt for fysikk, NTNU, Trondheim, Norway}
\affiliation{D.~V.~Skobeltsyn Institute of Nuclear Physics,
Moscow State University, Russia}

\author{R.~Tom\`as}
\affiliation{II.~Institut f\"ur theoretische Physik,
Universit\"at Hamburg, Germany}

\date{March 12, 2009}

\begin{abstract}
High energy photons from blazars interact within tens of kpc with the
extragalactic photon background, initiating electromagnetic pair cascades.
The charged component of such cascades is deflected by extragalactic
magnetic fields (EGMF), leading to halos even around initially point-like
sources.
We calculate the intensity profile of the resulting secondary high-energy
photons for different assumptions on the initial source spectrum and the
strength of the EGMF, employing also fields found earlier in a constrained
simulation of structure formation including MHD processes.
We find that the observation of halos around blazars like Mrk~180
probes an interesting range of EGMF strengths and acceleration models: 
In particular, blazar halos test if the photon energy spectrum at the source 
extends beyond $\sim 100$\,TeV and how anisotropic this high energy 
component is emitted.
\end{abstract}

\pacs{
98.70.Sa,    
95.85.Pw,    
98.54.Cm,    
98.62.En     
}

\maketitle

\section{Introduction}
\label{intro}

Observations of photons with energies up to tens of TeV, mainly
by imaging air Cherenkov telescopes (IACT), have opened in the last
decade a new window on the non-thermal side of the Universe~\cite{review}.
One of the main aims of these observations is the identification of the
sources of cosmic rays  and their acceleration mechanism. Observations
of very high energy photons can however provide information on many
more areas of astrophysics, as e.g.\ the amount of extragalactic background
light (EBL) and the strength of extragalactic magnetic fields (EGMF).

The tight relationship between the EBL, the strength of EGMFs and the horizon
of high energy photons is well-known: Photons with energy  above the pair 
creation threshold, $E_{\rm th}\sim30$\,TeV for $E\sim 10^{-2}$\,eV as 
characteristic energy of the background photons,
can interact with them, 
initiating
electromagnetic pair cascades~\cite{1}. The charged component of such
cascades is in turn deflected by magnetic fields, first by  fields in the
cluster surrounding the source~\cite{Aharonian:1993vz} and then by
EGMFs~\cite{Neronov:2007zz}. If the EGMF is not extremely week,
$B\lsim 10^{-16}$\,G, secondary photons are deflected outside
a point-like source and thus the observed  point-like photon flux  $I(E)$ 
consists only of the surviving primary photons, 
$I(E)=\exp[-\tau_{\gamma\gamma}(E)]I_0$, with $\tau_{\gamma\gamma}$ as the 
depth for pair production. 
If the EGMF exceeds $10^{-10}$\,G in a large volume fraction, the halo
becomes too large and its intensity  therefore too small for a detection.
In the intermediate range, ($10^{-16}$--$10^{-10}$)\,G,
the deflections and thus the halo extension may be sufficiently small
that an observation by current or next generation IACTs may be
feasible~\cite{Neronov:2007zz}.

We focus in this work on the effect of EGMFs on electromagnetic cascades,
providing detailed predictions for the expected intensity $I(E,\theta)$  of
the halo produced by cascades as function of the observed energy $E$
and the angular distance $\theta$ to the source. Since the EGMF is
highly structured and the Milky Way is contained in the supergalactic
plane, the magnetic field along the line-of-sight to different blazars,
and thus also their halos, will vary significantly, even if they
would be at the same distance and would have the same source energy spectrum.
It is therefore crucial to use an EGMF model that describes at least
qualitative correctly the location of voids and filaments. Since deflections
close to the observer are most important, it is sufficient to account for
the EGMF structure of the local universe.
For our predictions of the expected halo intensity $I(E,\theta)$ around
selected blazars we use therefore EGMF models calculated in Ref.~\cite{weak}
that are based on a constrained simulation of the large-scale structure
within $\approx 115$\,Mpc.

Our results confirm the estimate of Ref.~\cite{Neronov:2007zz} that
blazar halos are detectable for specific acceleration models, 
if EGMFs are as weak as found in
the simulations of Ref.~\cite{weak}. Thus blazar halos may provide valuable
information about the strength of EGMFs in a range where other tools like
Faraday rotation measurements can give only upper limits. Moreover,
our detailed predictions for the  expected intensity $I(E,\theta)$ of the
halo produced by the electromagnetic cascade
allow one to disentangle the contributions from
secondaries generated close to the source,
as suggested in Ref.~\cite{Aharonian:1993vz},
and during propagation in the EGMF.
The intensity of the halo is also sensitive to the total luminosity of
the source above $\sim 100$\,TeV and to the opening angle $\theta$ of the
cone in which most of this high energy radiation is emitted.

We find that the measured energy spectrum of most
blazars detected with IACTs may be described by a broken power-law that is
steeper in the measured energy range (100\,GeV--few~TeV) than usually assumed,
if one adds a softer high energy component.  This additional high-energy
component is absent both in Synchrotron Self-Compton SSC and Inverse Compton
models for blazars~\cite{Sikora:2001ze}, but may be caused by
hadrons~\cite{proton,cena}. In most hadronic  models, e.g.\ acceleration
close to the core in electromagnetic fields or in hot spots, the bulk of
radiation, although emitted anisotropically,  extends over a cone with
opening angles $\alpha_{\rm jet}\gsim 10^\circ$. For such conditions, 
we found that
the halo is not strongly suppressed. Thus the observation of blazar halos
would be an indication for an extension of the source photon spectrum
beyond 100\,TeV that is not strongly beamed and thus also evidence for the
acceleration of hadrons in blazars.

This article is structured as follows: We start in Sec.~\ref{scheme} with
a brief description of our Monte Carlo (MC) scheme. Then we study in
Sec.~\ref{toy}
a toy model that allows us to exhibit clearly the dependence of
the halo properties on external parameters as the strength of the EGMF,
the source photon energy spectrum and the amount of beaming. In
Sec.~\ref{sources}, we use
magnetic fields from Ref.~\cite{weak} along the line-of-sight towards four
selected blazars and calculate their halos, choosing the source
energy spectrum such that the observed photon spectra at low energies
are reproduced. Finally, we summarize in Sec.~\ref{sum}.

\section{Monte Carlo scheme}
\label{scheme}

We choose to describe the propagation of the electromagnetic (e/m)
cascade in the
extragalactic space using a MC method. Starting from an initial photon emitted
by the source, we trace the development of the cascade taking into account
$e^+e^-$ pair production by photons and inverse Compton scattering (ICS) of
electrons due to their interactions with photons from the radio, infra-red,
and the microwave background as well as synchrotron energy losses of electrons.
For the spectral densities of the backgrounds we use the ``Best-Fit06'' model
from Ref.~\cite{ir} for the EBL and Ref.~\cite{radio} for the radio background.

\begin{figure}[htb]
\begin{center}
\epsfig{file=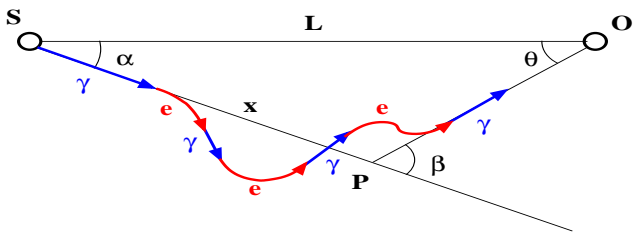,height=3 cm,width=.47\textwidth}
\end{center}
\caption{
Schematic view of the electromagnetic cascade in extragalactic space.
\label{fig:prop}}
\end{figure}

Since we are interested in gamma-rays arriving to the observer with
small angular deflections, we can use an essentially one-dimensional picture
for the  propagation of the e/m cascade.
A typical propagation pattern is sketched in Fig.~\ref{fig:prop}.
The initial photon is emitted by the source $S$ with the angle $\alpha$ with
respect to the direction to the observer $O$. Due to interactions of both the
initial photon and of secondary leptons, an e/m cascade develops over the
distance $x$, until at point $P$ the final photon is produced. 
Note that in the ultrarelativistic limit relevant for us, the daughter
particles move in the forward direction.
The final photon
arrives to the observer under the angle $\theta$ with respect to the direction
to the source, while its wave-vector forms the angle $\beta$ with respect to the
initial photon direction. The three angles $\theta$, $\alpha$,
and $\beta$ are related by
\begin{eqnarray}
\alpha &=& \beta - \theta \,,\\
  \sin \theta & =& \frac{x}{L}\, \sin \beta \,.
\end{eqnarray}

For small $\theta$ we are interested in, we have
\begin{equation}
 \theta \simeq \frac{x}{L}\, \sin \beta \,.\label{theta}
\end{equation}

We calculate the average scattering angle $\beta$ within the small angle
approximation as
\begin{equation}
\beta =\sqrt{\langle \beta ^2\rangle}=\sqrt{\sum_i  \beta_i^2},
\end{equation}
with $\beta_i^2$ as  angular deflection of the electron $i$ during the
cascade development. The deflection angle in a regular magnetic field is 
\be
  \beta_i\simeq\frac{d}{R_L}\simeq 0.52^\circ
  \left(\frac{p_\perp}{10^{20}\,{\rm eV}}\right)^{-1}
  \left(\frac{d}{1\,{\rm kpc}}\right)
  \left(\frac{B}{\mu{\rm G}}\right)
\ee
for a particle with momentum $p_\perp$ perpendicular to $B$ propagating
the distance $d$.

As it can be read off from Eq.~(\ref{theta}), the effect of deflections $\beta$
at large distances from the observer is suppressed by the ratio  $x/L$
compared to those close by. Hence an important contribution
to the halo at small $\theta$ are final photons produced  at $x\ll L$, and
passing the remained distance $L-x$ without interactions.
A competing although less important contribution comes
from photons with small scattering angle $\beta$. In particular, the
latter is somewhat enhanced in the case of a structured  EGMF: Filaments of
the EGMF may be crossed by intermediate photons in the e/m cascade, while
intermediate electrons propagate in the voids.
As a result, the strength and the structure of the magnetic field close
to the observer is more important than the ``average'' field.

\begin{figure} 
\epsfig{file=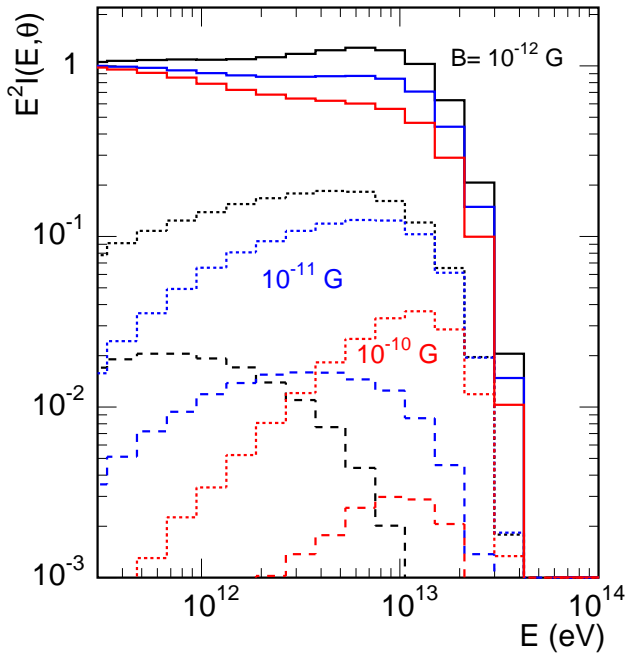,width=0.47\textwidth}
\caption{
The differential energy spectrum $I(E,\theta)$ for three
different angular bins and a uniform magnetic field with strength
$B=10^{-12}$\,G (black), $10^{-11}$\,G (blue) and $B=10^{-10}$\,G (red).
\label{fig:toyBl}}
\end{figure}

\section{Parameter dependence of the halo in a toy model}
\label{toy}

\begin{figure*} 
\begin{tabular}{cc}
\epsfig{file=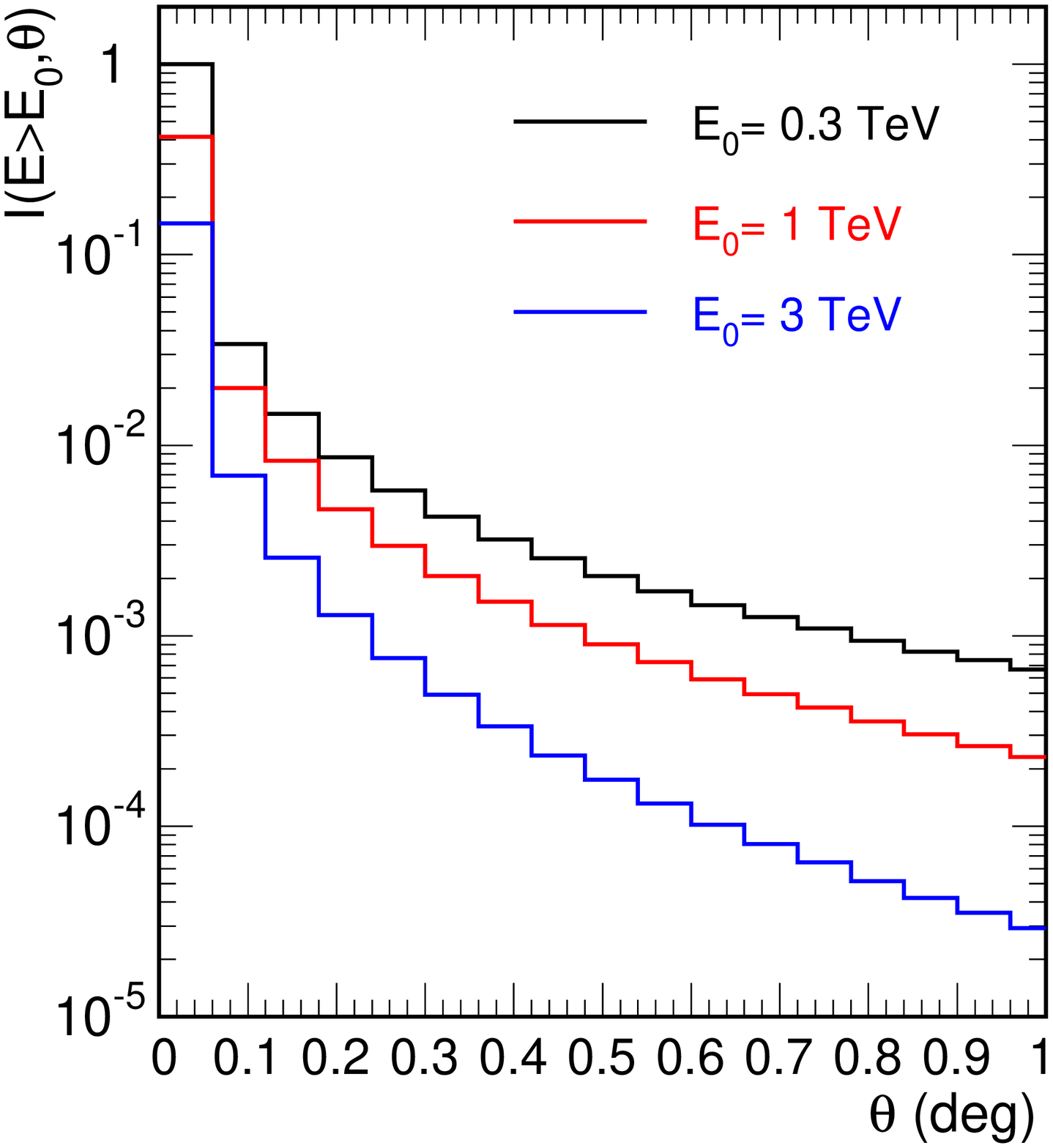,width=0.47\textwidth,height=0.45\textwidth} &
\epsfig{file=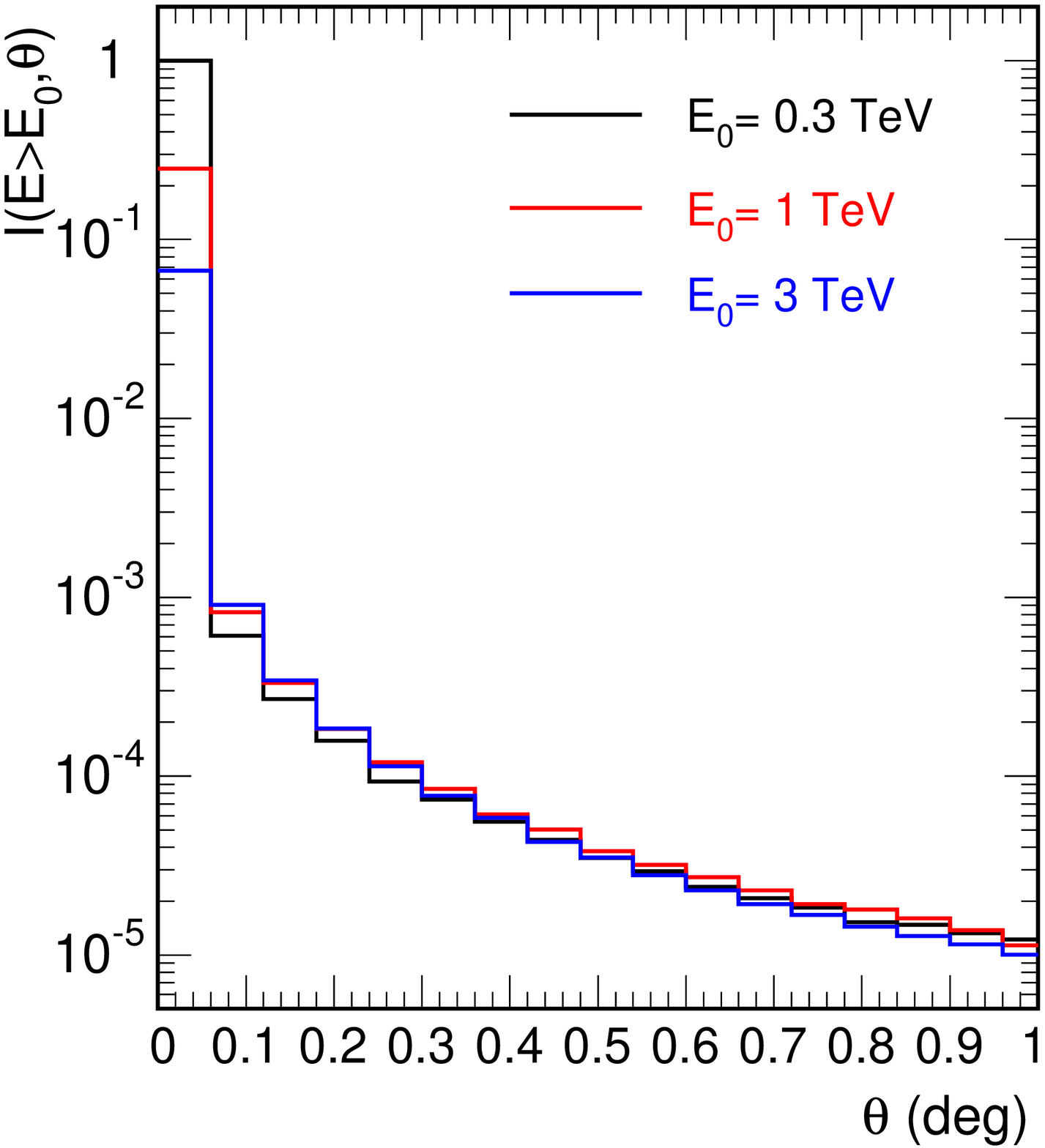,width=0.47\textwidth,height=0.45\textwidth}\\
\end{tabular}
\caption{
The integral intensity $I(>E_0,\theta)$ as function of the angular distance
$\theta$ to the source for a uniform magnetic field with strength
$B=10^{-12}$\,G (left) and $B=10^{-10}$\,G (right).
\label{fig:toyBr}}
\end{figure*}

The intensity and the extension of the halo around a blazar caused
by the deflection of $e^+e^-$ pairs in the electromagnetic cascade
depends mainly on the strength and the structure of the EGMF,
the spectral shape of the source spectrum and the total luminosity of the
source above $\sim 30$\,TeV. In order to disentangle better the
influence of these different factors, we employ in this section a toy model:
We use a uniform extragalactic magnetic field and a simple power-law
as the energy spectrum of photons leaving a blazar.

If not otherwise specified, we use as reference parameters a uniform EGMF
with field strength $B=10^{-12}$\,G, a power-law for the source
energy spectrum of photons, $dN/dE\propto E^{-\gamma}$, with $\gamma=2$,
$E_{\max}=10^{17}$\,eV and place our reference source at the distance of
135\,Mpc, i.e.\ representative for Mrk~501.

\paragraph{Dependence on the EGMF strength}

We first discuss the change of the intensity profile
$I(E,\theta)=d^2N(E,\theta)/d\Omega$ varying the strength of the EGMF.
Figure~\ref{fig:toyBl} shows the differential energy spectrum $I(E,\theta)$ as
function of the observed energy $E$ for three $\theta$ bins: 
The solid lines include both the point-like and the halo  component
within $\theta<\theta_0=0.06^\circ$ (from now on called 'point-like'), 
while the dotted lines denote the mean intensity
between $\theta_0<\theta<2\theta_0$ ('halo 1') and
the dashed lines denote the intensity at $\theta=1.2^\circ$
('halo~2').
Our choice for the angular binning is related to the angular accuracy of
present IACT (typically, few arcmin) and to experimental methods
of studying blazar halos: The intensity of $\gamma$-radiation
from the angular vicinity of the source is compared to the background
intensity in the field of view of the apparatus (typically within 1 degree).
Thus, the feasibility of halo measurements can be characterized by
the relative difference
in the integral intensity between  'halo~1' and 'halo~2'. On the other hand,
the overall sensitivity of an instrument to the halo component can be
determined from the intensity ratio between the 'point-like' and 'halo~1'
components.   
As the strength of the EGMF increases from $B=10^{-12}$\,G (left)
to $B=10^{-10}$\,G (right), the  $I(>E_0,\theta)$
distributions shown in Fig.~\ref{fig:toyBr} flatten, since the deflection
angles increase. Moreover, the  $I(>E_0,\theta)$ distributions for
different $E_0$ become more similar to each other.
As a result, the ratio between the halo intensity at small $\theta$ and
the point-like intensity decreases too.
As one can see from the figures, even for relatively strong magnetic fields
with $B\sim 10^{-10}$\,G a halo observation is possible, in principle,
if one restricts oneself with gamma energies in excess of 10 TeV.
However, imposing such a cutoff would result in a drastic reduction of the
observed flux, the measurement thus being hardly possible with present IACT
sensitivity.
Thus, the observation of blazar halos
is more likely, if the line-of-sight to the blazar is contained in a void,
where the fields of strength $B=10^{-12}$\,G may be typical.

\begin{figure*}[t]
\begin{center}
\begin{tabular}{cc}
\epsfig{file=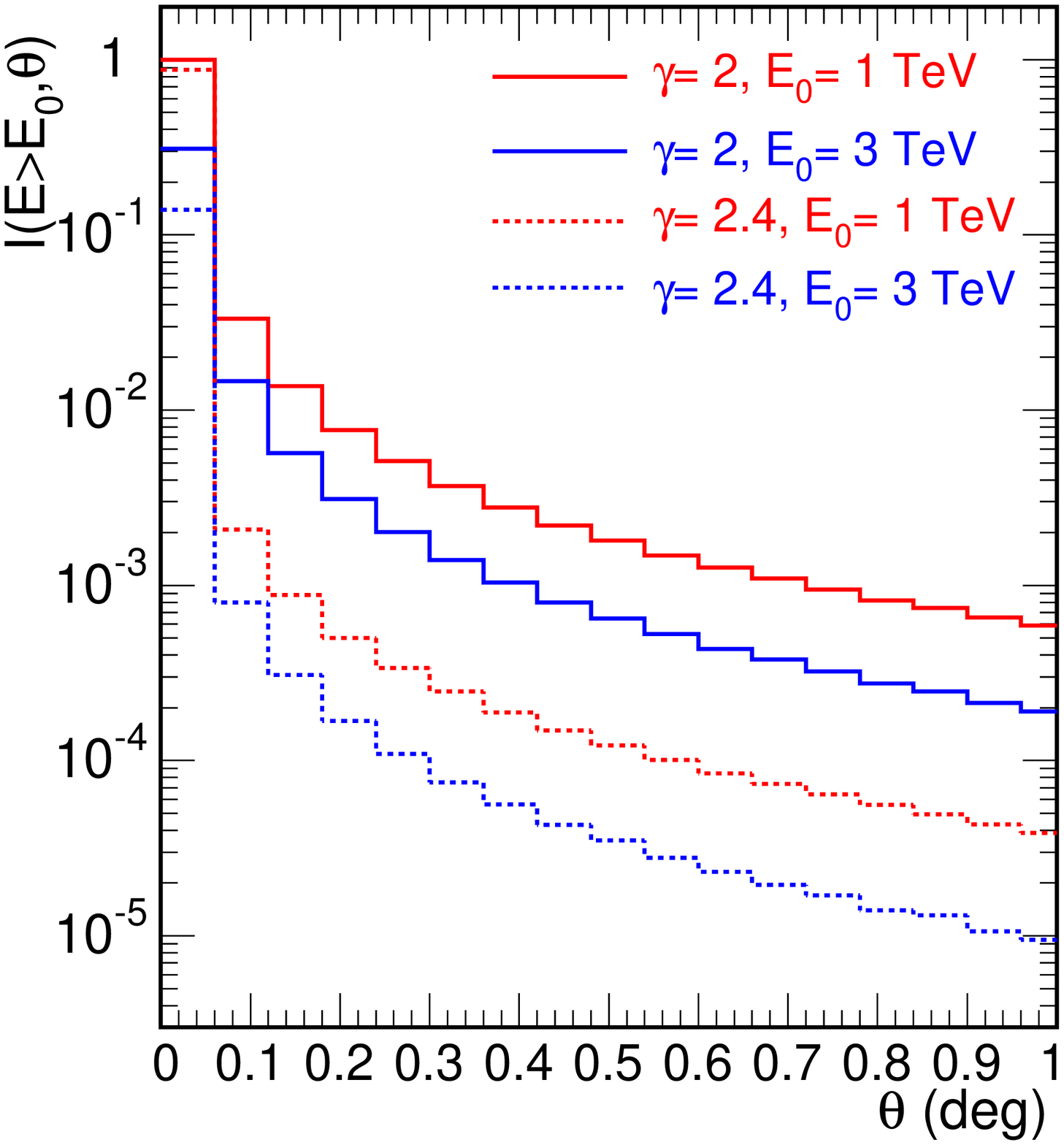,width=0.47\textwidth,height=0.47\textwidth} &
\epsfig{file=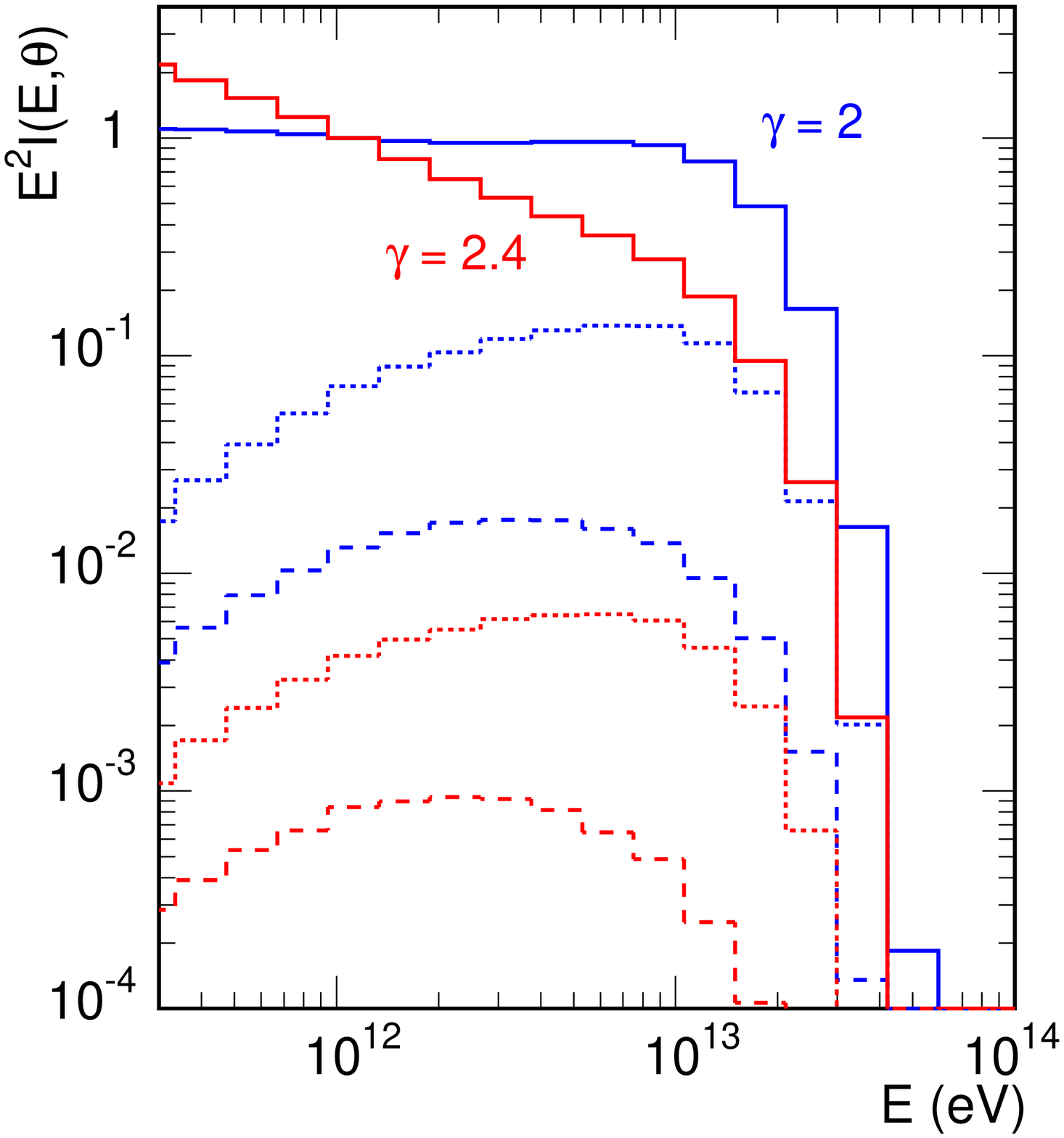,width=0.47\textwidth,height=0.47\textwidth}\\
\end{tabular}
\end{center}
\caption{
Comparison of $\theta$-dependence of integral intensity $I(>E_0,\theta)$
(left panel) and of differential energy spectra $I(E,\theta)$ (right panel)
for different source spectra:
power-laws  with $\gamma=2.0$ and  $\gamma=2.4$ (lower curves
in the right panel; solid lines for point-like, dotted for 'halo 1',
and dashed for 'halo 2').
A uniform magnetic field with strength $B=10^{-11}$\,G and the distance 
134\,Mpc is assumed.
\label{fig:toyg}}
\end{figure*}

\paragraph{Dependence on the spectral slope}
A comparison of the intensity profiles for two different values
of the exponent, $\gamma=2.0$ and 2.4, in the power-law for the source
energy spectrum of photons, $dN/dE\propto E^{-\gamma}$, is shown in
Fig.~\ref{fig:toyg}. The halo component depends more weakly on the spectral
slope $\gamma$ than the point-like component, but its relative
normalization decreases sharply for steeper source energy spectra. 
As a result,
the ratio of halo and point flux decreases strongly towards lower energies.
Nevertheless, the total number of halo events increases lowering the
threshold energy $E_0$. The optimal energy to look for the
halo component depends also on the background of cosmic ray events and
the background of point-like events with misreconstructed arrival direction.
Therefore detailed studies taking into account the specific detector
properties are required for such an optimization.

\begin{figure}
\begin{center}
\epsfig{file=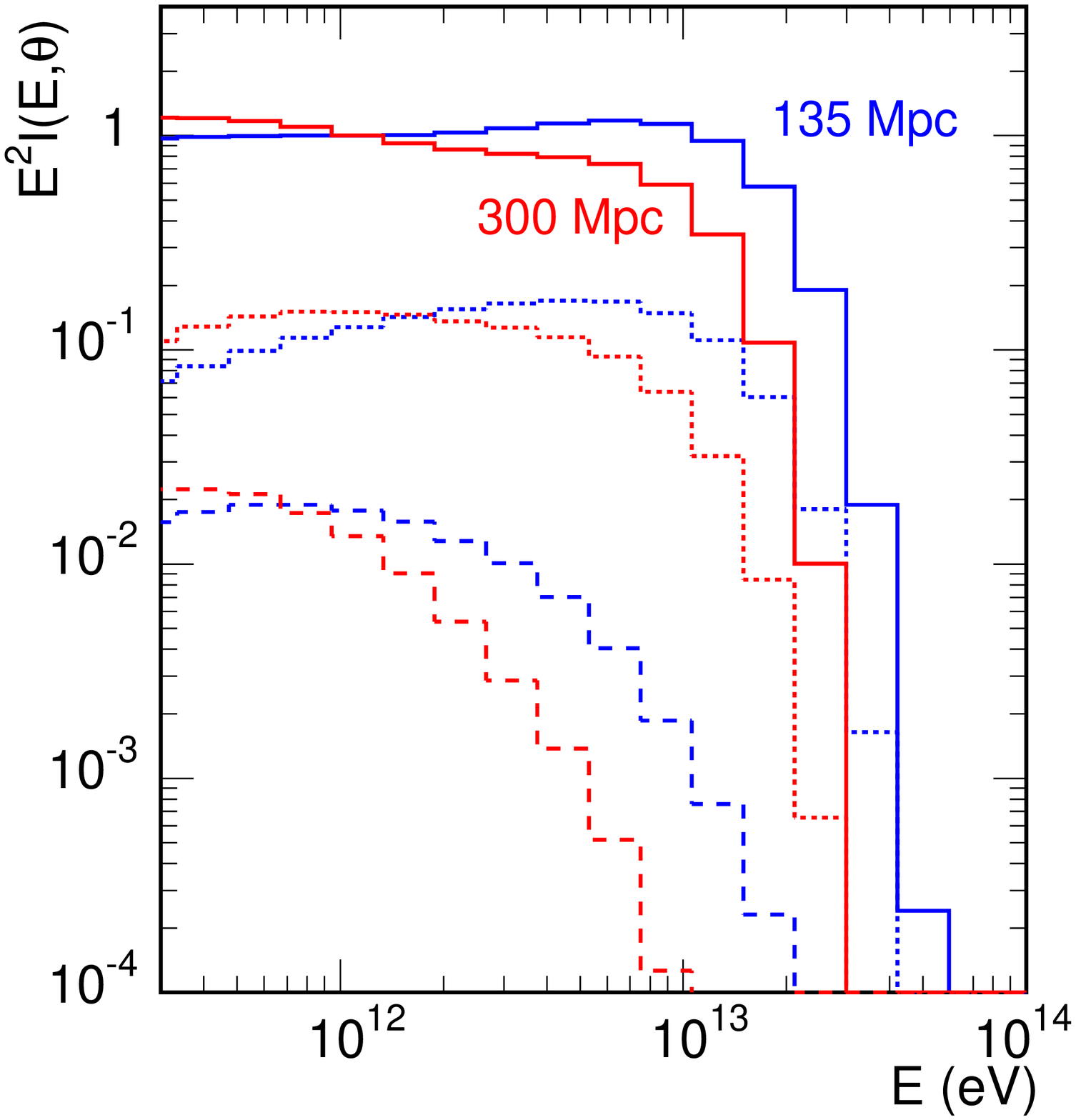,width=0.47\textwidth,height=0.45\textwidth}
\end{center}
\caption{
Comparison of differential energy spectra for three angular bins
for 135 and 300 Mpc source distance.
\label{fig:toyd}}
\end{figure}

\paragraph{Dependence on the  distance}
We compare the differential energy spectrum $I(E,\theta)$ of the point-like
to the 'halo 1' and 'halo 2' components for two different distances 
in Fig.~\ref{fig:toyd}. If one factors out the trivial
distance dependence, i.e.\ normalizes the flux at say 1\,TeV to a constant
value, then the halo components of sources at 135 and 300\,Mpc show
only minor differences: This behavior is explained by the fact
that deflections at the distance $l$ to the observer are weighted with 
the factor $\sim (L-l)/L$. Hence,
deflections close to the observer contribute most to the size of
the halo, while interactions at large distances lead mainly to
a reduction of the observed flux.  Moreover, the strongest deflection
in case of a uniform EGMF considered here comes from the propagation
of the last electron in the e/m cascade, which has the smallest energy,
hence, is most strongly deflected by the field.

\begin{figure}
\begin{center}
\epsfig{file=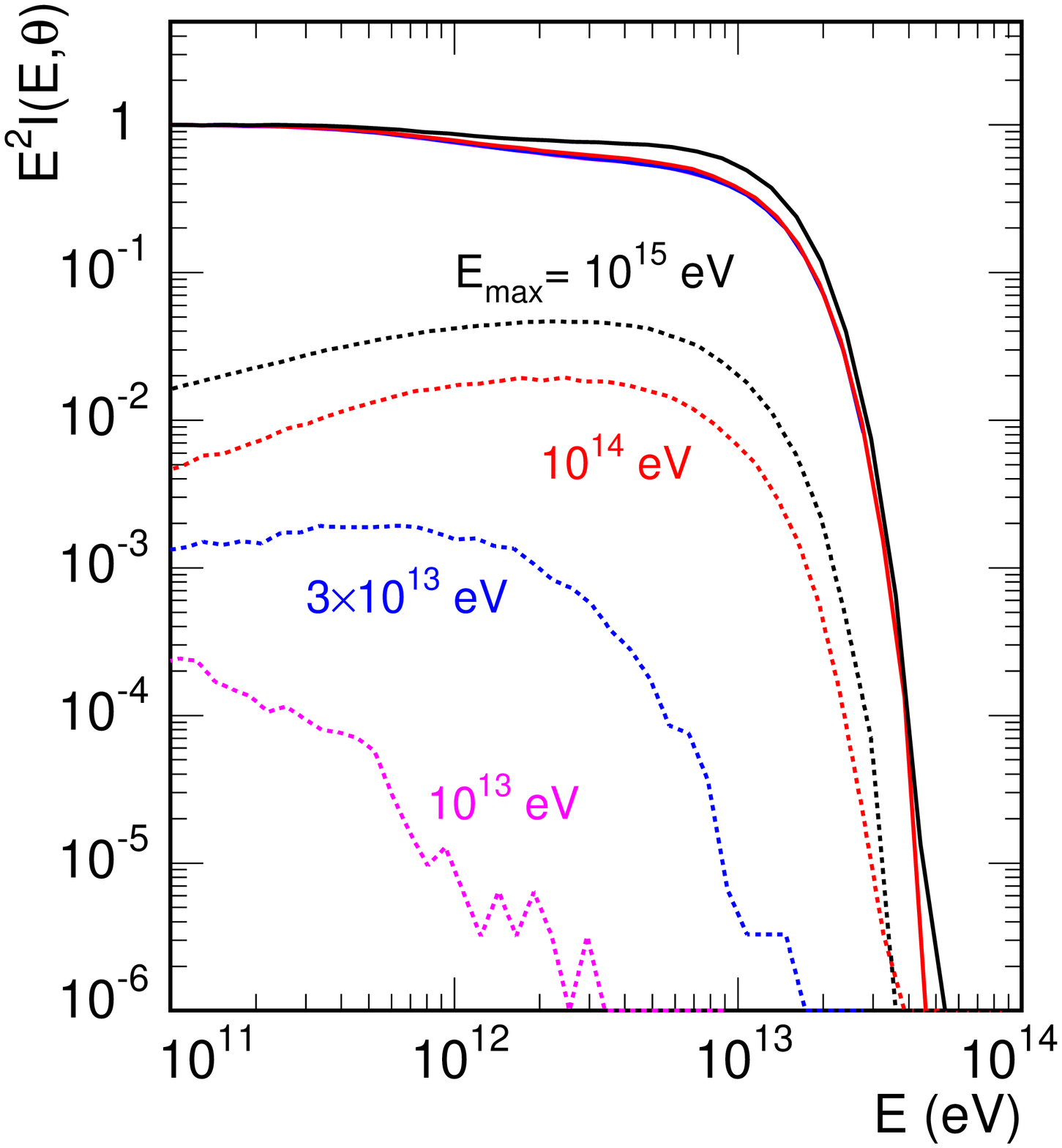,width=0.47\textwidth,height=0.45\textwidth}
\end{center}
\caption{
Comparison of the differential energy spectra for the point-like (three solid
lines at the top) and ``halo 1'' (dashed lines) components, 
with three different energy cutoffs of the source spectrum:
top down $E_{\max}=10^{15},10^{14}, 3\times 10^{13}$ and $10^{13}$\,eV.
\label{fig:toyEmax}}
\end{figure}

\paragraph{Dependence on high-energy luminosity}
We use again a power-law $dN/dE\propto E^{-2}$, varying now the maximal
energy $E_{\max}$ from $E_{\max}=10^{13}$, $3\times 10^{13}$, $10^{14}$ to
$10^{15}$\,eV.  In Fig.~\ref{fig:toyEmax}, the influence of the luminosity
injected above the  pair-production threshold on the halo intensity is
clearly visible. While close to the threshold an increase of $E_{\max}$
leads to a strong enhancement of the halo component, a further increase of
$E_{\max}$ beyond $10^{15}$\,eV has  much smaller influence of the
halo intensity for the chosen $1/E^2$ spectrum.

\paragraph{Dependence on the jet opening angle $\alpha_{\max}$}
Finally we discuss the influence on the halo intensity of an anisotropic
emission of the high energy photons. Qualitatively, one expects from
Fig.~1, that restricting  $\alpha$ to smaller and smaller values
for a given $\theta$
requires that the last interaction at $P$ happens closer and closer
to the observer. Thus the intensity of the halo will be reduced on all
angular scales, but we do not expect an angular cutoff as effect of an
anisotropic emission by the source.

\begin{figure*}[t]
\begin{center}
\begin{tabular}{cc}
\epsfig{file=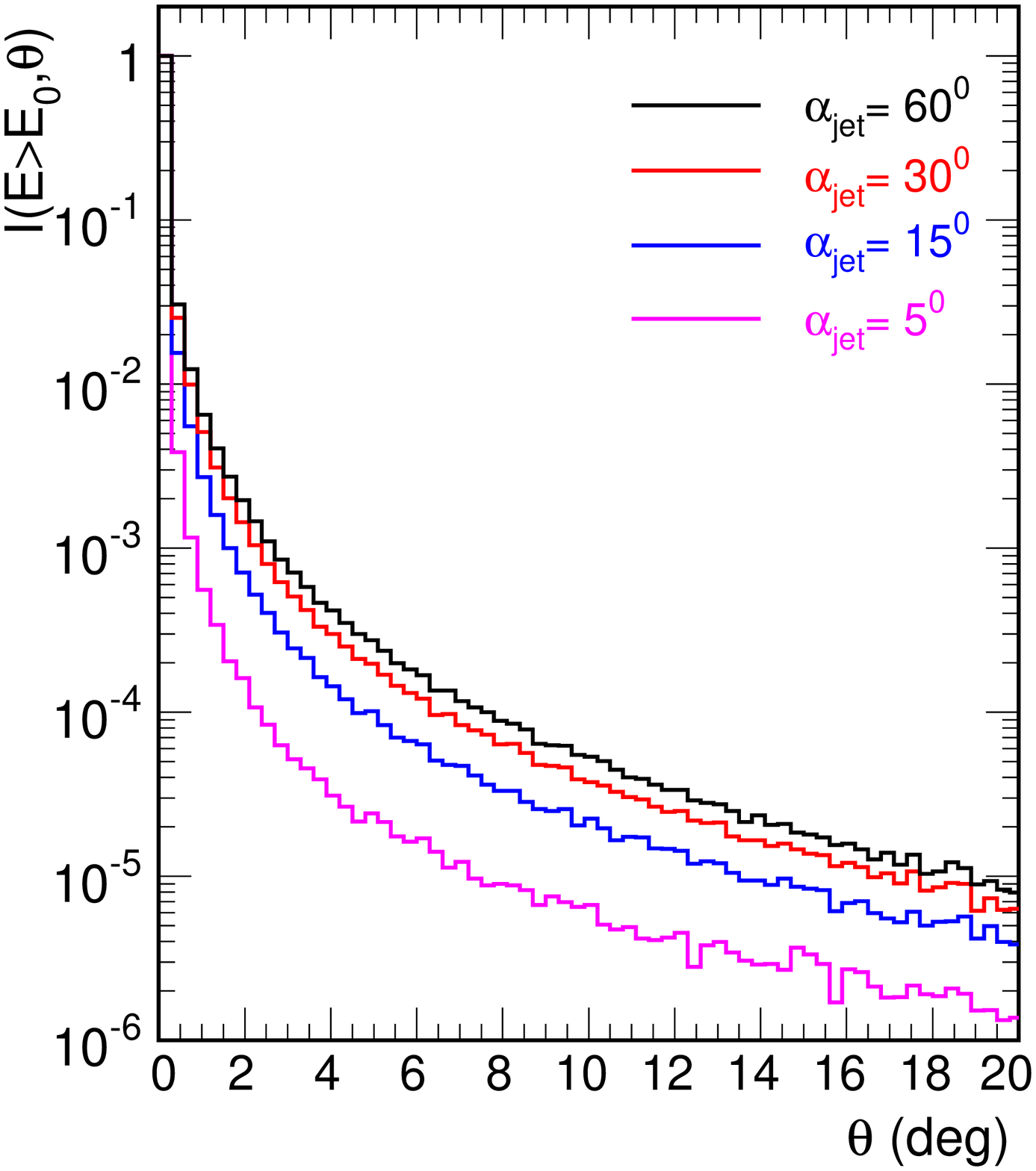,width=0.47\textwidth,height=0.47\textwidth} &
\epsfig{file=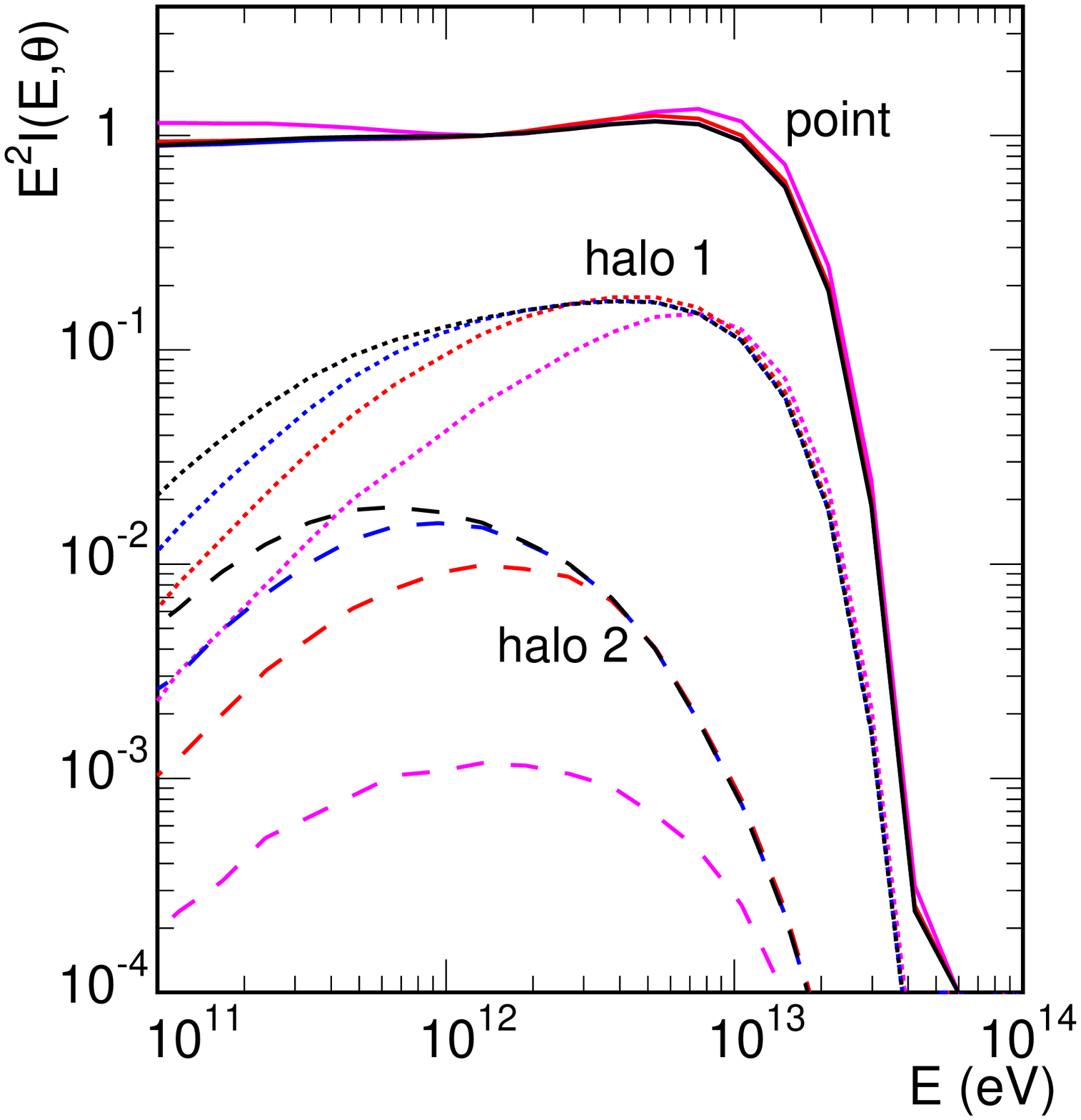,width=0.47\textwidth,height=0.47\textwidth}\\
\end{tabular}
\end{center}
\caption{
Comparison of the $\theta$-dependence of the integral intensity 
$I(>E_0,\theta)$ (left panel) and of the differential energy spectra 
$I(E,\theta)$ (right panel) for four different  jet opening angles, 
$\alpha_{\rm jet}=60^\circ,30^\circ, 15^\circ$ and $5^\circ$ (top down).
\label{fig:toyjet}}
\end{figure*}

In order to obtain a quantitative understanding for the importance
of beamed emission, we have recalculated our standard case 
(isotropic gamma-ray emission in the source) restricting
possible emission angles $\alpha$ with respect to the line-of-sight
to the blazar to $\alpha_{\rm jet}=60^\circ,30^\circ,15^\circ$ and $5^\circ$.
The resulting integral intensity $I(\theta,>300{\rm GeV})$ as function of
$\theta$ is shown in the left panel of Fig.~\ref{fig:toyjet}, while we
show the differential energy spectrum $I(E,\theta)$ of the point-like
to the 'halo 1' and 'halo 2' components in the right panel. The
integral intensity $I(\theta,>300{\rm GeV})$ shows clearly an overall
suppression as function of $\theta$ instead of a cutoff. The extent
of this suppression becomes smaller at higher energies, but clearly
the detection of blazar halos would disfavour the emission of the
high energy component in a very narrow cone, $\alpha_{\rm jet}\lsim 10^\circ$.


In summary, we have found that the shape of the halo depends sensitively
on all investigated parameters, but generally $I(E,\theta)$ decreases
faster than the $1/\theta$ behavior suggested in Ref.~\cite{Neronov:2007zz}.
The obtained  enhancement of the small $\theta$ region, compared to
Ref.~\cite{Neronov:2007zz}, is due to the precise MC description of the
cascade process, being related to the
contributions from the tails of the distribution of the electron's 
interaction length and of the secondary photon energies in ICS: Shorter
distances travelled by
electrons and higher energy of the parent electron producing the final
photon result in smaller than average values of $\theta$. Similarly, 
decreasing the opening angle of the high energy emission of photons
in the source does not produce an angular cutoff in the observed photon
spectrum, contrary to what has been suggested by the authors of
Ref.~\cite{Neronov:2007zz}. The latter was an artifact of considering
only the first two interaction steps in the electromagnetic cascade
in Ref.~\cite{Neronov:2007zz}.
Most importantly, we found that  the intensity of the halo can exceed
1\% of the point-like contribution at energies in excess of few TeV
for field strengths $B= 10^{-12}$\,G.

\section{Halo predictions for selected blazars}
\label{sources}


\begin{figure*}[!tbp]
\begin{center}
\begin{tabular}{cc}
\epsfig{file=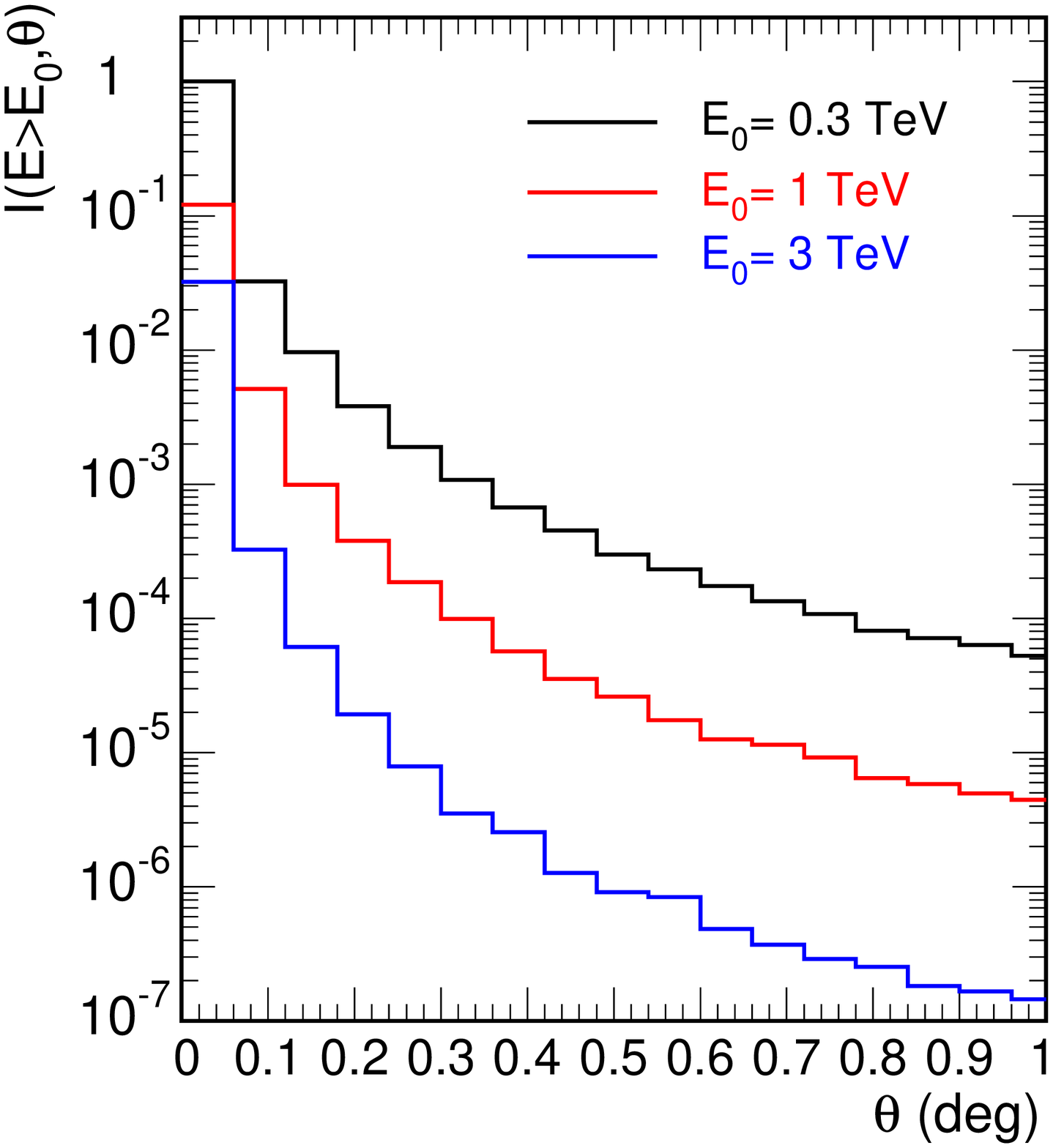,width=0.47\textwidth,height=0.45\textwidth} &
\epsfig{file=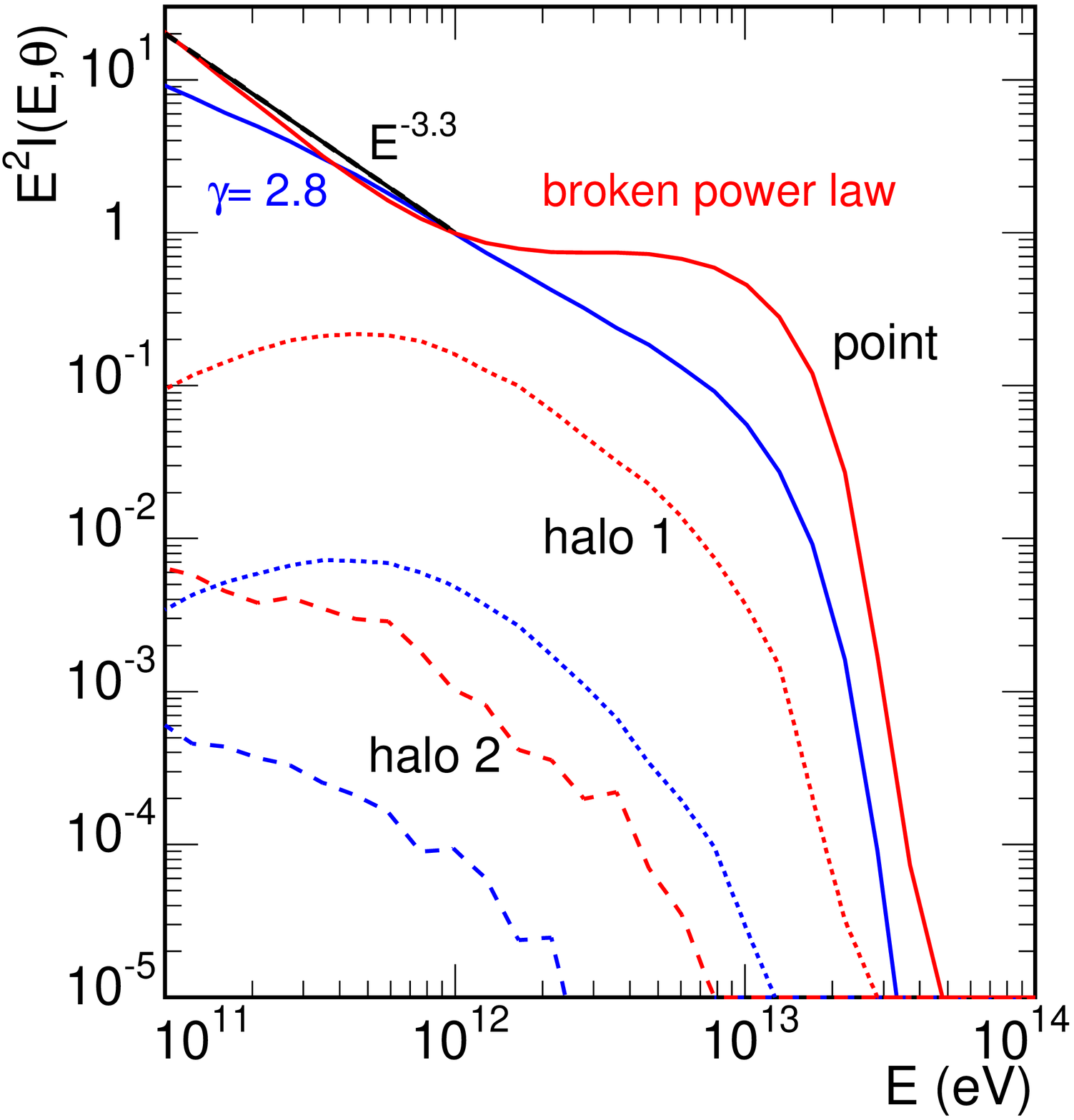,width=0.47\textwidth,height=0.45\textwidth}\\
\end{tabular}
\begin{tabular}{cc}
\epsfig{file=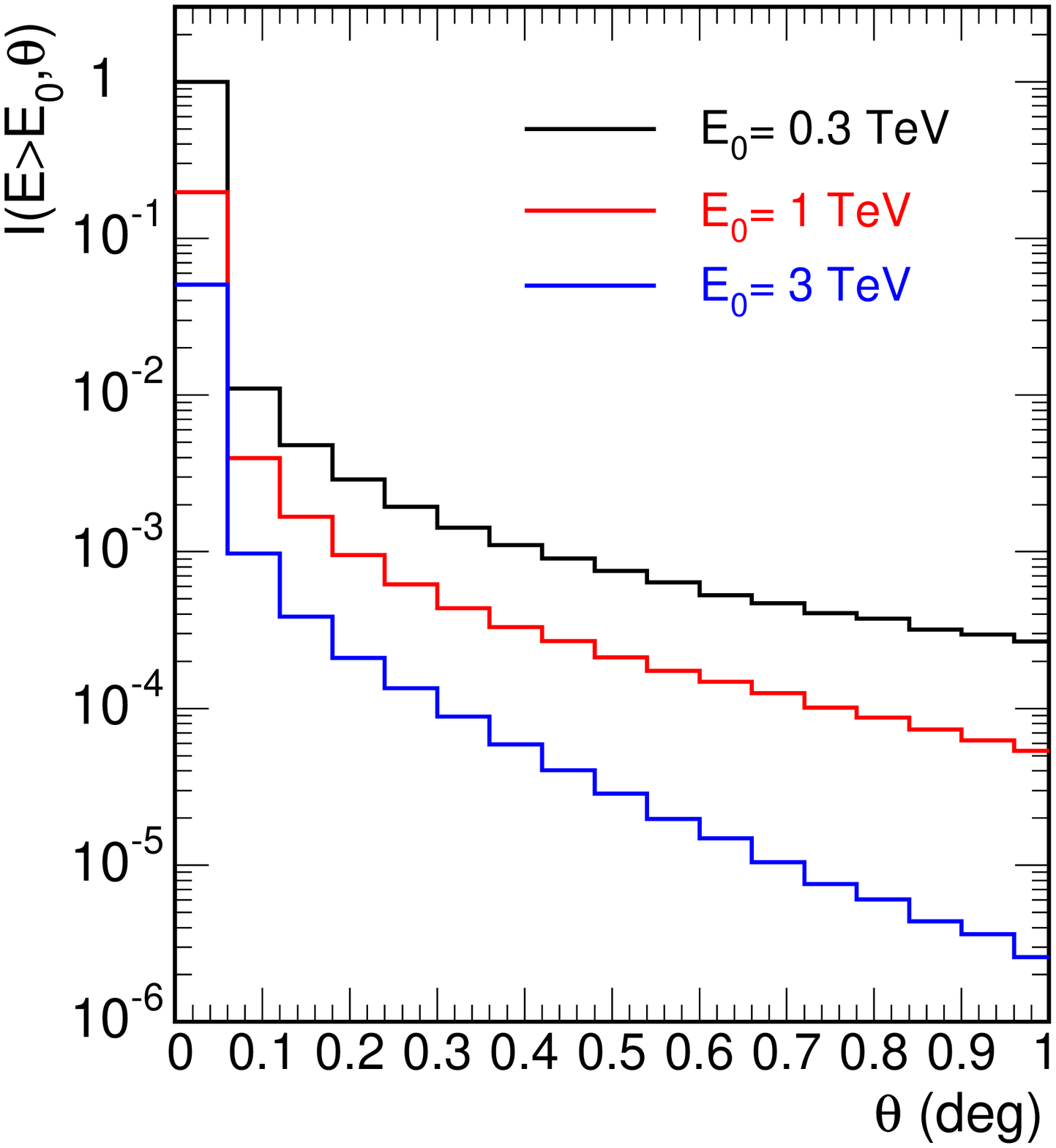,width=0.47\textwidth,height=0.45\textwidth} &
\epsfig{file=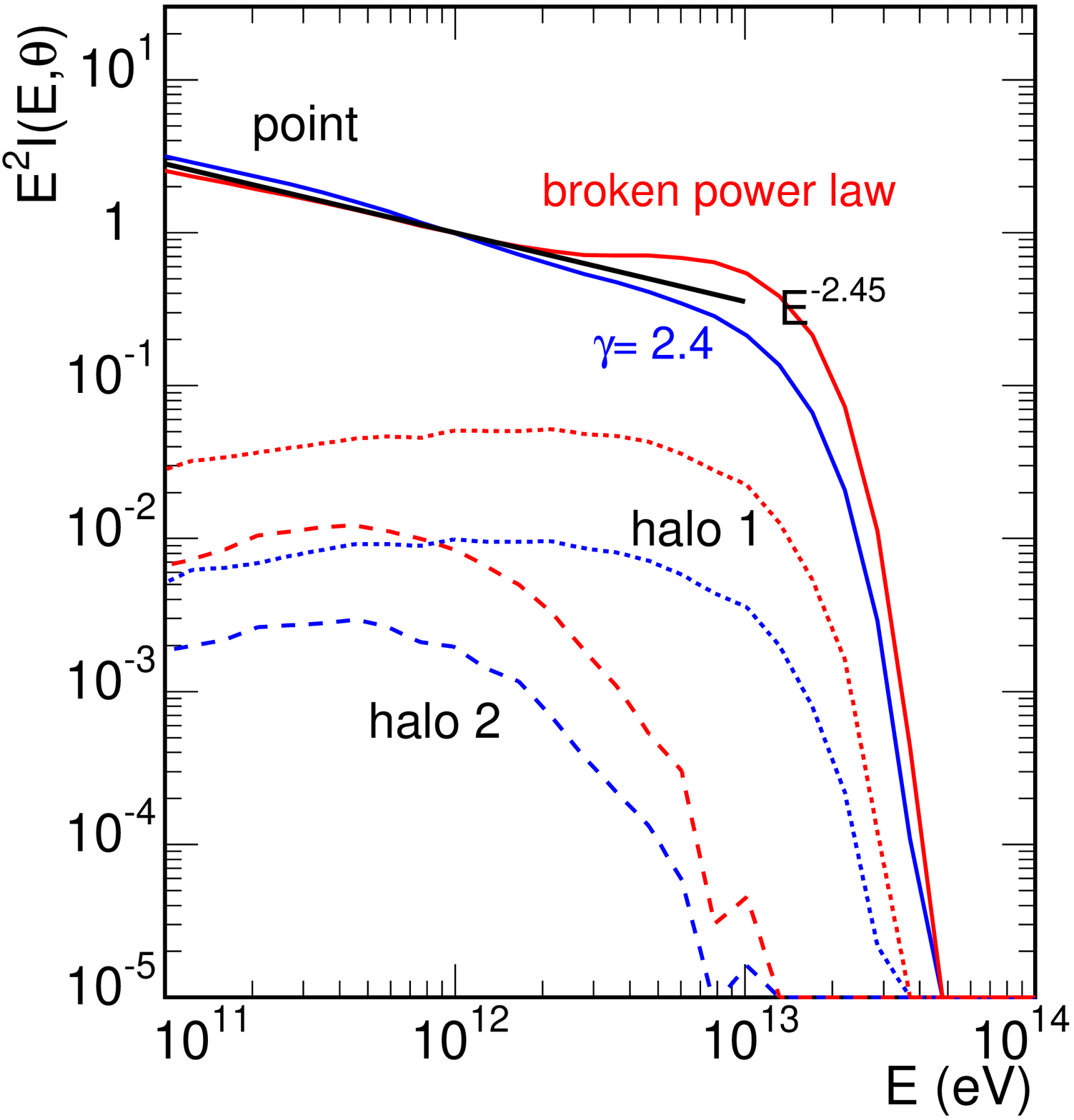,width=0.47\textwidth,height=0.45\textwidth}\\
\end{tabular}
\end{center}
\caption{
Top: Mrk 180 (left panel: $\theta$-dependence of integral intensity for the
``broken'' power-law initial spectrum with
$\gamma_1=3.5$, $\gamma=1.8$ and $E_b=2\,$TeV;
right panel: differential energy spectra for three angular bins
for the power-law source spectrum with $\gamma=2.8$ and for the
``broken'' power-law case),
bottom: Mrk 501 (left panel: $\theta$-dependence of integral intensity for the
``broken'' power-law initial spectrum with
$\gamma_1=2.4$, $\gamma=2.0$ and $ E_b=5\,$TeV;
right panel: differential energy spectra for three angular bins
for the power-law source spectrum with $\gamma=2.4$ and for the
``broken'' power-law case);
for all cases $E_{\max}=10^{17}\,$eV.
\label{fig:reala}}
\end{figure*}

\begin{figure*}[!tbp]
\begin{center}
\begin{tabular}{cc}
\epsfig{file=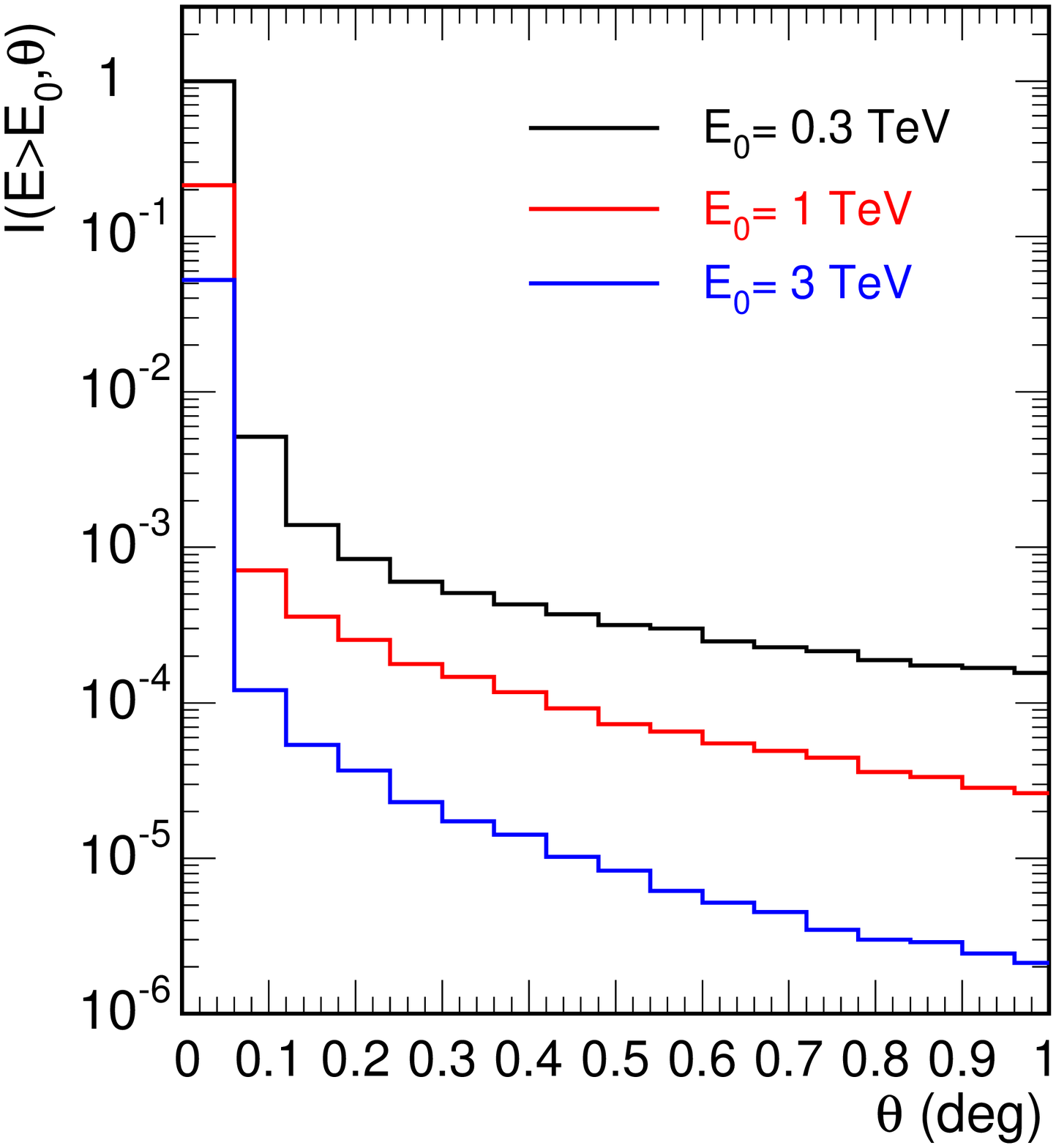,width=0.47\textwidth,height=0.45\textwidth} &
\epsfig{file=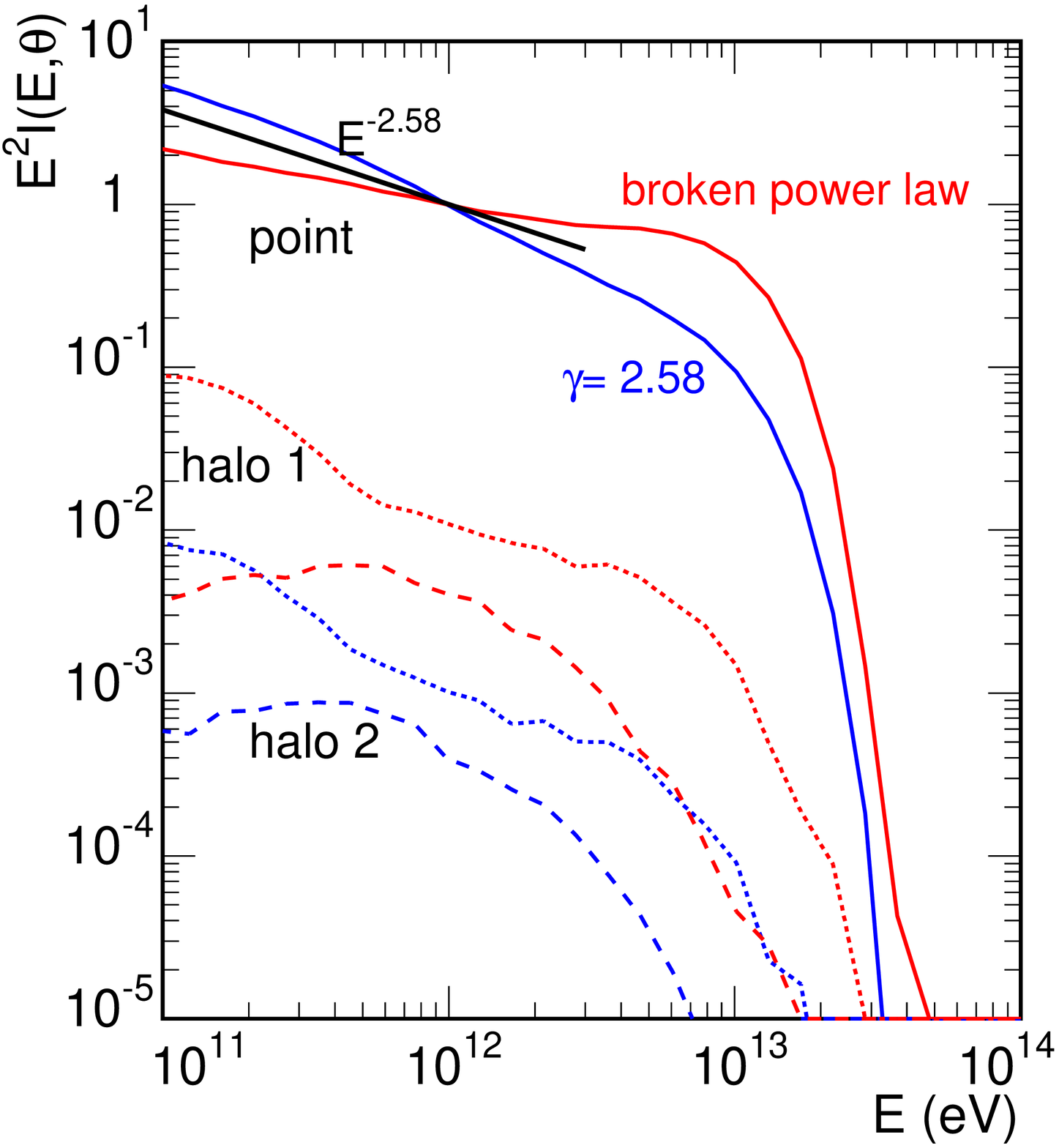,width=0.47\textwidth,height=0.45\textwidth}\\
\end{tabular}
\begin{tabular}{cc}
\epsfig{file=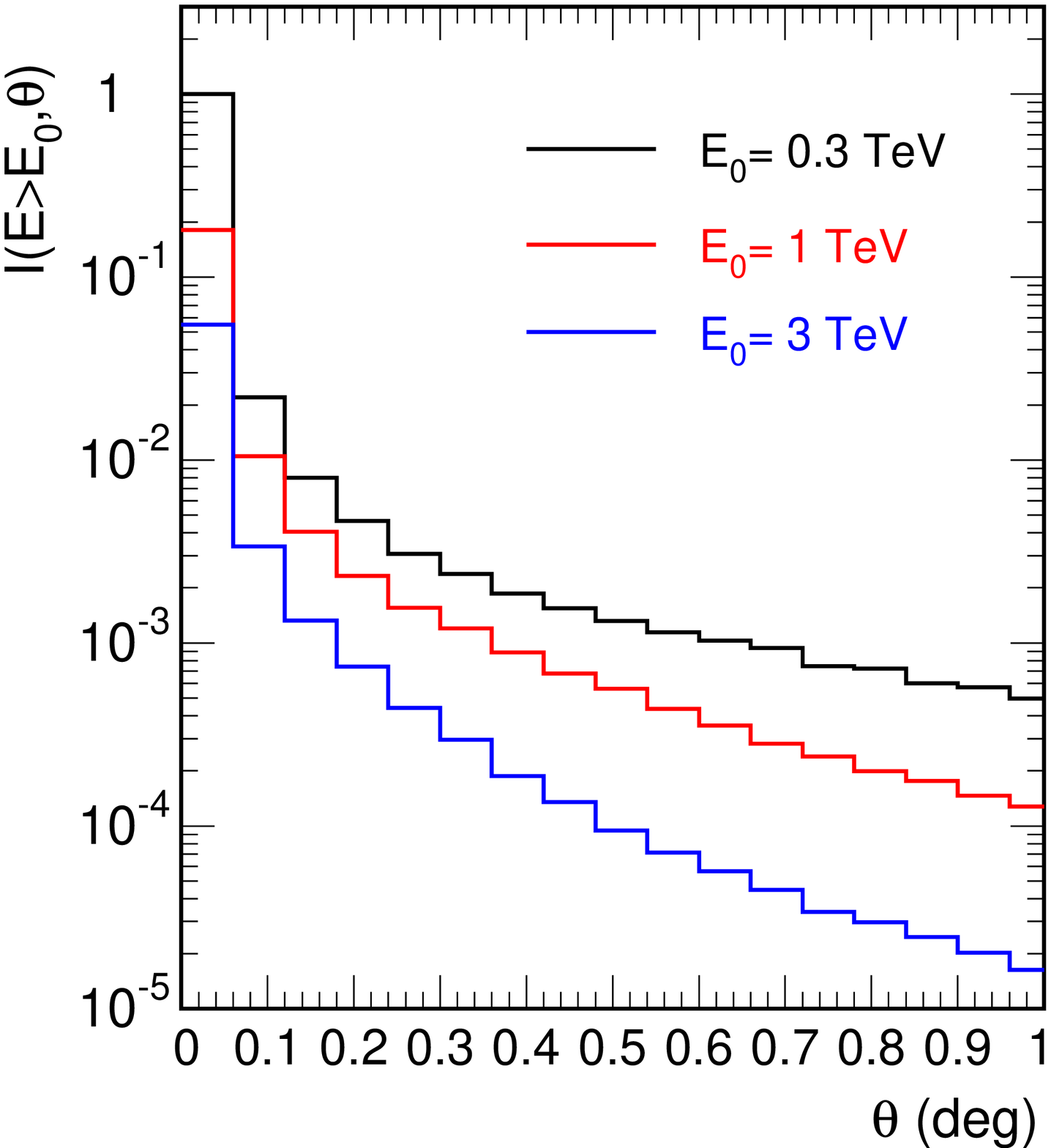,width=0.47\textwidth,height=0.45\textwidth} &
\epsfig{file=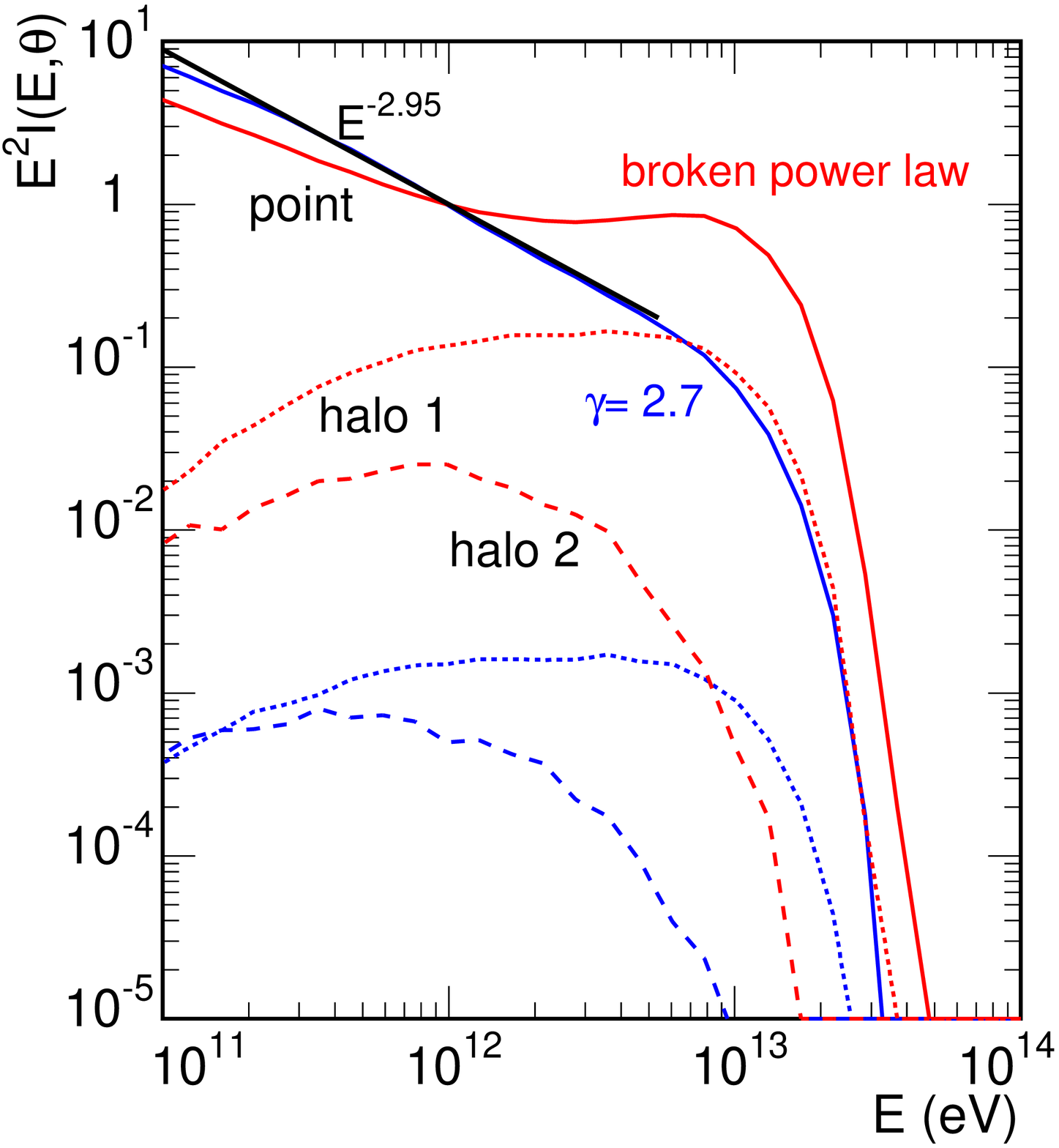,width=0.47\textwidth,height=0.45\textwidth}\\
\end{tabular}
\end{center}
\caption{
top: 1ES1959+650, (left panel:  $\theta$-dependence of integral
intensity for the ``broken'' power-law initial spectrum with
$\gamma_1=2.6$, $\gamma=1.8$ and $E_b=3\,$TeV;
right panel: differential energy spectra for three angular bins
for the power-law source spectrum with $\gamma=2.58$ and for the
``broken'' power-law case),
bottom: 1ES2344+514 (left panel:  $\theta$-dependence of integral
intensity for the ``broken'' power-law initial spectrum with
$\gamma_1=2.7$, $\gamma=1.7$
and $E_b=5\,$TeV;
right panel: differential energy spectra for three angular bins
for the power-law source spectrum with $\gamma=2.7$ and for the
``broken'' power-law case).
\label{fig:realb}}
\end{figure*}

We now turn from the toy model studied in the last section to
a realistic study of the blazar halos for the five nearest blazars.
The table~\ref{tab:sources} summarizes the distance, the observed
($\gamma$) and intrinsic ($\gamma_s$) spectral index assuming a power-law,
the energy of the observed cut-off or the energy of the last measured
data point as well as the reference used. As far as possible we used the
photon spectra  measured during quiet periods. While in the case of
Mrk~421 the maximal energy $E_{\max}=1.4$\,TeV corresponds to an
observed cutoff in the measured energy spectrum, the numerical
value of $E_{\max}$ corresponds otherwise to the last data point given
in the cited references. From our discussion of the toy model, it is
clear that the low luminosity of Mrk~421 above the pair creation threshold
makes it an unfavorable candidate for the search of blazar halo. Therefore
we will restrict our analysis to the remaining four blazars,
Mrk~180, Mrk~501, 1ES~1959+650 and 1ES~2344+514.

\begin{figure}[!tbp]
\begin{center}
\epsfig{file=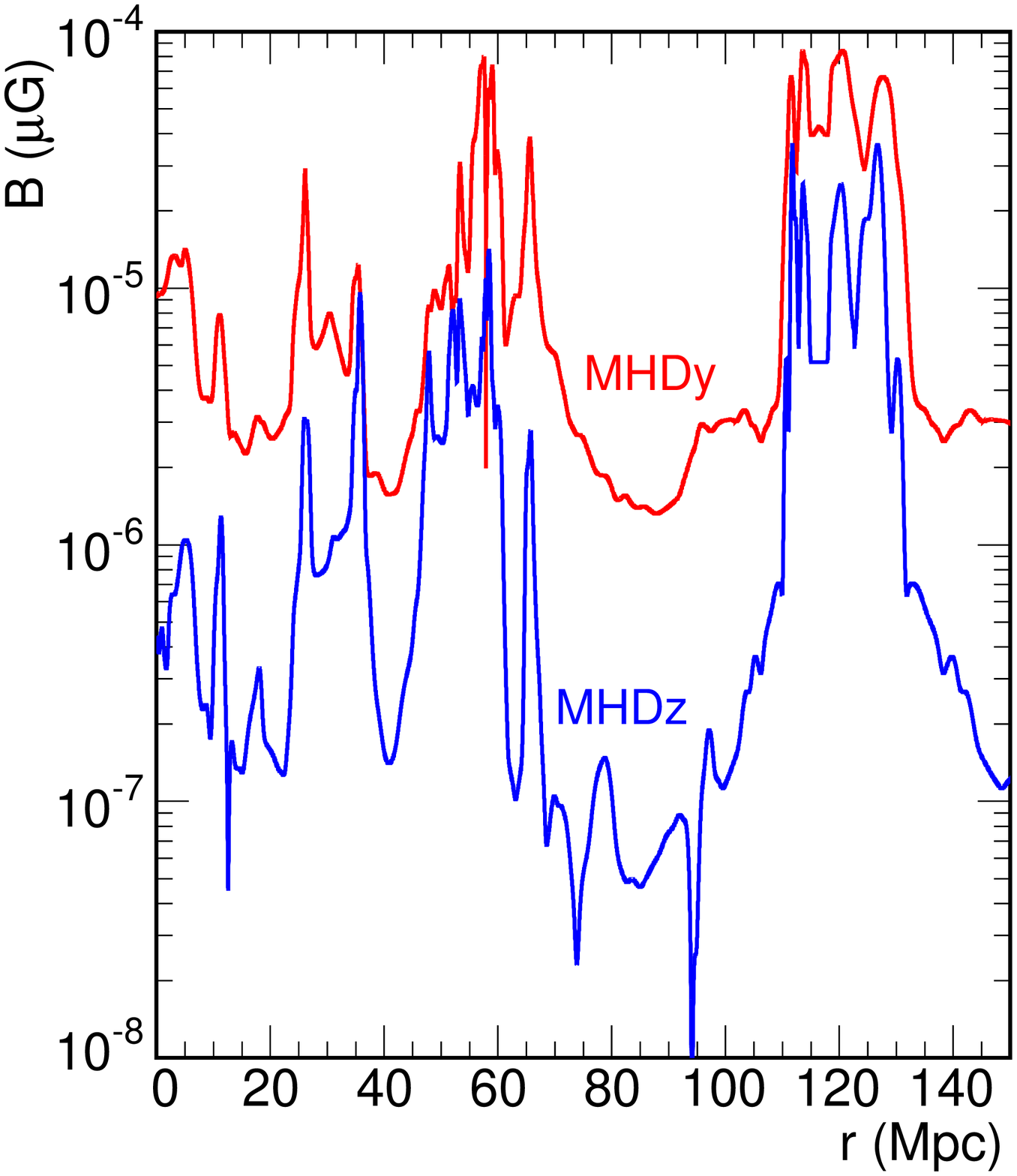,width=0.47\textwidth,height=0.4\textwidth}
\end{center}
\caption{
The magnetic field $B_\perp$ perpendicular
to the line-of-sight towards Mrk~180.
\label{fig:B}}
\end{figure}

We used  the EGMF along the line-of-sight towards the four blazars
that results from a constrained realization of the local universe
(see Ref.~\cite{weak} and references therein). In short,
the initial density fluctuations were
constructed from the IRAS 1.2-Jy galaxy survey by smoothing the
observed galaxy density field on a scale of 7\,Mpc, evolving it
linearly back in time, and then using it as a Gaussian constraint
(see \cite{Hoffman1991}) for an otherwise random realization of the
$\Lambda$CDM cosmology. The volume constrained by the IRAS
observations covers a sphere of radius $\approx 115\,$Mpc centered on
the Milky Way and therefore encloses the prominent local structure
including a fair representation of the large scale density field.
Many of the most prominent clusters observed locally can therefore
be identified directly with halos in the simulation, and their
positions and masses agree well with their simulated counterparts
(see also \cite{2002MNRAS.333..739M}). Although the smoothing
of the observed galaxy density field when producing
the initial conditions invokes spacial uncertainties
within the simulated structures up to the used smoothing scale of 7\,Mpc,
the statistics of crossing filaments and voids within the local universe
should nevertheless be quite realistic. This is especially the case
for the objects considered in this study as we can make use of the
full path length through the simulated volume as these objects
are lying outside the simulated region.

For this work we used the magnetic field configuration obtained
from two realizations (MHDy and MHDz) starting from
different initial seed fields (for more details see Ref.~\cite{weak})
having lower (MHDz) and higher (MHDy) initial values for the magnetic
seed field. Both simulations lead to magnetic fields within the
galaxy clusters which are statistically in agreement with the still
rare Faraday rotation measures and are roughly bracketing
the allowed range of magnetic fields in galaxy clusters.

A large fraction of the line-of-sight towards the blazers of
interest are within the volume covered by this simulation,
however it can cover only up to distances of $\approx 115$\,Mpc
and we are forced to extend the magnetic field profile beyond that point.
Since it is most important to reproduce the local field structure, we
expect that the details of the chosen extension have only a minor
influence as long the extended field has the correct statistical
properties. In practise, we mirrored the field at the point of maximal
distance along the line-of-sight. A comparison of the
EGMF component perpendicular to the line-of-sight towards Mrk~180
obtained in the models is shown in Fig.~\ref{fig:B}.

We will first present results for the model MHDz that generally predicts
stronger localisation of magnetic fields in cluster and filaments and thus
smaller deflections than model MHDy. Later, we compare the differences 
between our results for the two models.

We show our results\footnote{Numerical tables of these results can be
obtained from MK.} in Fig.~\ref{fig:reala} for Mrk~180 (top) and
Mrk~501 (bottom) and in
Fig.~\ref{fig:realb} for 1ES~1959+650 (top) and 1ES~2344+514 (bottom). 
The differential energy spectrum of each source is calculated for
two different injection spectra: In the first case, we assume that the
observed spectrum extends with the same slope $\gamma_1$ up to
$E_{\max}=10^{17}$\,eV. In the second case, we add a new, harder component
with $\gamma_2<\gamma_1$ that may be generated in particular by hadrons.
Thus, we use as injection spectrum a broken power-law, adjusting the
parameters like the break energy $E_b$ such that we reproduce the
observed spectral shape while optimizing the halo flux.

A characteristic feature of the second case is a small bump
between the break energy and the exponential cutoff at $E_{\rm th}$
introduced by the EBL interactions. Clearly, blazars which show a cutoff
at energies below $E_{\rm EBL}$ can have only a minor additional hard
component compared to blazars which show either no cutoff or even a turn up.
As one can see from the Figures, for the ``broken'' power-law case we get
very pronounced halos in the TeV range  for Mrk 180 and 1ES~2344+514,
with the halo intensity being comparable to the one of the
point-like component.

\begin{table}
\begin{center}
\begin{tabular}{lccccc}
Source  & $d_L$/Mpc & $\gamma$ &$\gamma_s$ & $E_{\max}$/TeV & Ref. \\
Mrk~421 & 130 & 2.2 & 2.2 & 1.4 &\cite{Albert:2006jd} \\
Mrk~501 & 132& 2.45 & - & 10 & \cite{Albert:2007zd}\\
1ES~2344+514 & 183& 2.95 & 2.66 & 5.4 & \cite{Albert:2006km} \\   
Mrk~180 & 194 & 3.3 & 2.8 & 1 &\cite{Albert:2006bc} \\
1ES~1959+650 & 198 & 2.58 & - & 3 & \cite{Tagliaferri:2008qk,Aliu:2005jv,Aharonian:2003be}
\end{tabular}
\end{center}
\caption{
List of considered blazars.
\label{tab:sources}}
\end{table}

\begin{figure*}[!tbp]
\begin{center}
\begin{tabular}{cc}
\epsfig{file=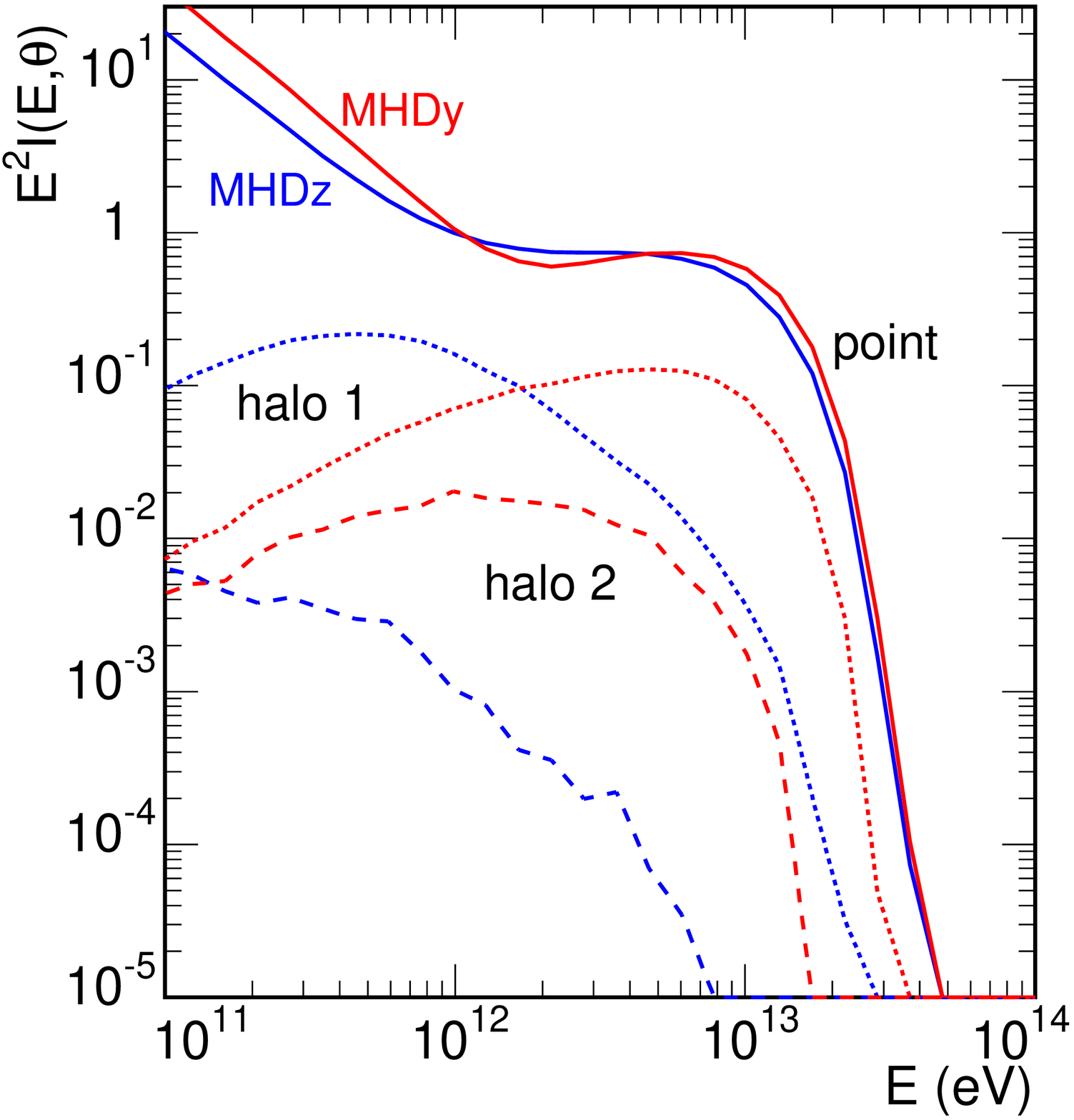,width=0.47\textwidth,height=0.45\textwidth} &
\epsfig{file=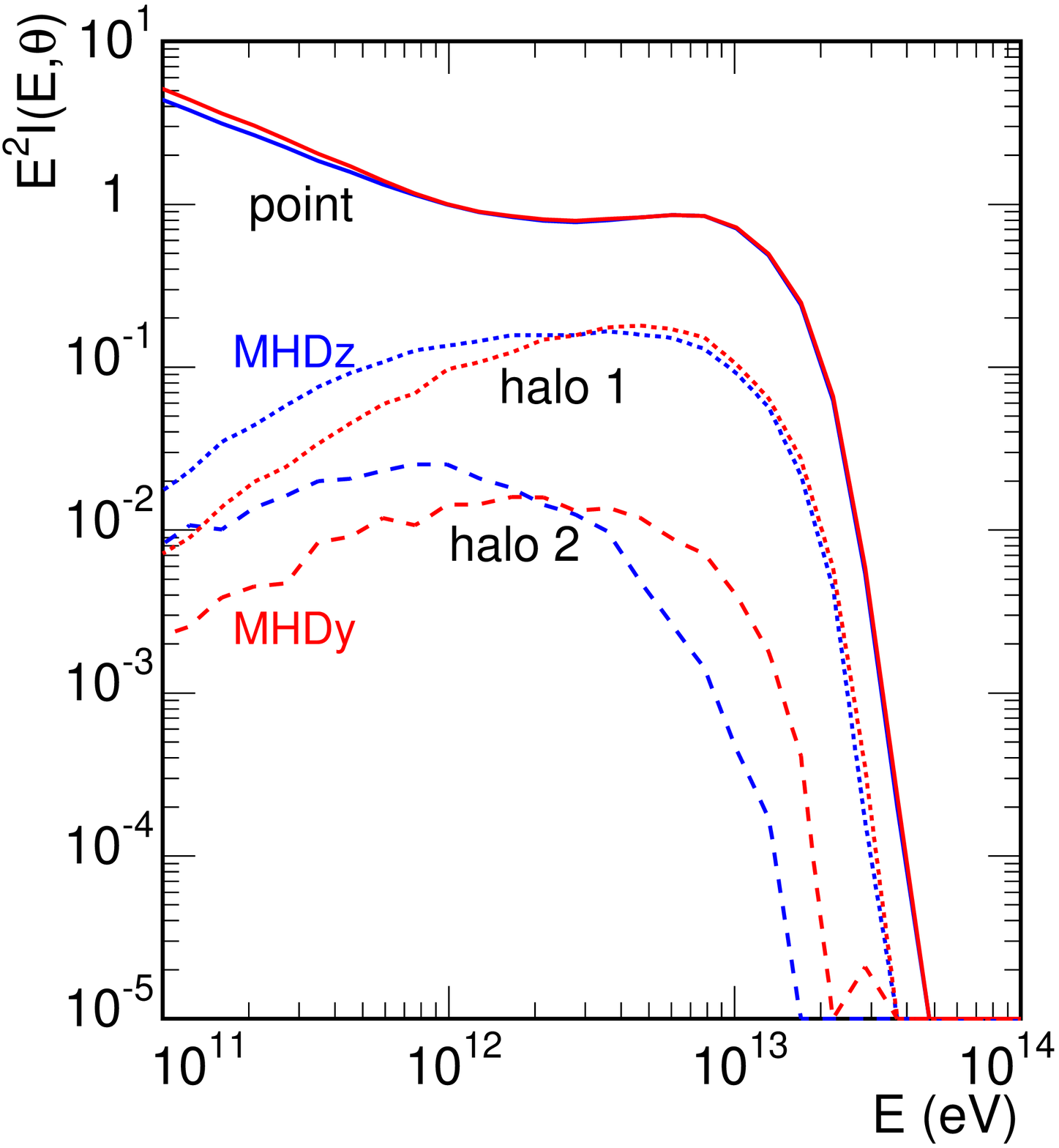,width=0.47\textwidth,height=0.45\textwidth}\\
\end{tabular}
\end{center}
\caption{
Comparison of differential energy spectra for three angular bins
for two realizations of EGMF (MHDy [red] and MHDz [blue]) for
Mrk~180 (left panel) and for  1ES~2344+514 (right panel).
\label{fig:yz}}
\end{figure*}

In Fig.~\ref{fig:yz}, we compare the energy spectrum of Mrk~180 (left)
and 1ES~2344+514 (right) obtained using the EGMF models MHDy and MHDz
together with the
``broken'' power-law initial spectra. For both sources, we obtain a reduction
of both point-like and halo intensities for the MHDy realization of EGMF which
is characterized by stronger and less localized magnetic fields.
As one can see from Fig.~\ref{fig:B}, the strength of the perpendicular
component of EGMF in the MHDy realization is almost an order of magnitude
higher than in the MHDz case. This results in a reduction of the
halo component for both sources considered, especially at lower energies.
Nevertheless, for both realizations of the magnetic field structure,
Mrk~180 and 1ES~2344+514 remain very promising cases for the experimental
detection of the halo in the TeV energy range.

Finally we note that the HEGRA experiment derived an upper limit of 1\%
on the halo flux from Mrk~521 at $0.5^\circ$ during a burst
period~\cite{Hegra}.

\section{Discussion}
\label{sum}

We have studied blazar halos generated by deflections of the charged
component of electromagnetic pair cascades.
Our analysis has accounted both for the kinematics of the pair-production
and Compton processes without approximations as well as for the local
structure of the EGMF by using EGMF models found earlier in a constrained
simulation of  structure formation including MHD processes.
The intensity of these blazar halos depends mainly on the strength and 
the structure of the EGMF:
Halos with detectable intensity require very weak EGMFs, comparable
to those found in the MHDz simulations of Ref.~\cite{weak}. The
direction of the EGMF vector close to the observer constitutes
an uncertainty that cannot be resolved by numerical simulations.
If the line-of-sight of a specific source and the EGMF vector are
aligned close to the observer, the effect of EGMFs will be reduced.

Another necessary condition for the observability of halos is the extension
of the source energy spectrum above $\sim 10^{14}$\,eV. Such a high 
energy extension of
the source spectrum may be generated by high energy protons
interacting with background photons in the source or emitting
synchrotron or curvature radiation. Note however that the acceleration 
of hadrons in the source does not necessarily lead to a pronounced ultrahigh 
energy tail of the accompanying gamma-ray spectrum. For instance, in the 
case of hadrons accelerated close to the core of the AGN, high energy 
gamma-rays start pair cascades on the background of ultraviolet photons. 
As a result, the spectrum of gamma-rays leaving the source is essentially 
cut-off at $10^{13}\,$eV, for an example see e.g.\~Ref.~\cite{cena}. 
Thus, the maximal energies assumed in the present work for the photons
leaving the source rather correspond to acceleration in
larger volumes, filled with less dense photon backgrounds and weaker
magnetic field than close to the AGN core. Typical examples for
such a scenario are the acceleration of protons in the outer or radio
jets of AGNs.

Throughout the paper, we implicitly assumed that the high energy part
of the source spectrum has a hadronic origin and, consequently, is not 
strongly beamed. Since we found that the halo intensity is strongly 
suppressed if this high energy component is emitted too in a very narrow 
cone, $\alpha_{\rm jet}\lsim 10^\circ$, an observation of blazar halos 
would indicate a different emission mechanism for the source energy 
spectrum above $\sim 10^{14}$\,eV, pointing to a hadronic origin. 
The obtained integral intensity $I(\theta,>300{\rm GeV})$ in the case
of a beamed UHE emission shows clearly an overall suppression as function 
of the observation angle, without angular cutoff. The extent
of this suppression becomes smaller at higher energies, but clearly
the detection of blazar halos would disfavour the beamed character
of  high energy photon emission.

The observed spectra of TeV blazar show strong time-variability. In contrast,
any intrinsic time-variability of the halo component is washed out by 
time-delays induced by the curved trajectories of the charged secondaries 
in the EGMF. Since the halo component has an hadronic origin in our scenario,
we do not expect even an intrinsic time-variability to start with.
Consequently, we used spectral data measured during 'quiet' periods for the 
sources considered. Similarly, experiments should compare the relative 
normalization of the halo to the 'pointlike' component during quiet periods.

In the most optimistic cases, the halo component found reaches 10\%
of the point-like component and may be already  detected or excluded
by current IACTs.
An observation of blazar halos would provide valuable information
on the EGMF along the line-of-sight to the source and on the
minimal luminosity  emitted above the pair creation threshold.
Moreover, a halo detection would point to an unbeamed emission of the
high-energy part of the source photon spectrum, providing evidence
for a different acceleration and emission mechanism than the
standard leptonic one.

Finally, we comment on the generation of halos not by the electromagnetic
cascade in the extragalactic space but close to the source. A calculation
of the halo intensity from this alternative mechanism requires a detailed
model for the photon backgrounds and the magnetic field in the source.
Moreover, the model for acceleration, in particular the acceleration site,
has to be specified. While it is thus clear that an additional contribution
to the halo from processes close to the source exists, detailed predictions
will be strongly model dependent.

\section*{Acknowledgments}
We would like to thank Tanja Kneiske, Andrii Neronov, Julian Sitarek and 
especially Dieter Horns for helpful comments and suggestions.
S.O.\  acknowledges a Marie Curie IEF fellowship from the European Community,
R.T \  partial support from the Deutsche Forschungsgemeinschaft within the
SFB 676.


\end{document}